\documentclass[graybox]{svmult}

\usepackage{type1cm}        
\usepackage{makeidx}         
\usepackage{graphicx}        
\usepackage{multicol}        
\usepackage[bottom]{footmisc}
\usepackage{newtxtext}       %
\usepackage{newtxmath}       

\makeindex             

\newcommand{\R}{\ensuremath{{\mathbb R}}}
\newcommand{\N}{\ensuremath{{\mathbb N}}}
\newcommand{\Z}{\ensuremath{{\mathbb Z}}}
\newcommand{\CC}{{\mathcal C}}
\newcommand{\PP}{{\mathcal P}}

\newcommand{\MSE}{\mathsf{MaxError}}
\newcommand{\Encm}{\mathsf{Enc2}}
\newcommand{\Enc}{\mathsf{Enc}}
\newcommand{\Dec}{\mathsf{Dec}}

\begin{document}

	\title*{Comprehensive Introduction to Fully Homomorphic Encryption for Dynamic Feedback Controller via LWE-based Cryptosystem}
	\titlerunning{LWE-based Cryptosystem and Dynamic Controller for Dummies}
	\author{Junsoo Kim, Hyungbo Shim, and Kyoohyung Han}
	\institute{Junsoo Kim and Hyungbo Shim \at ASRI, Dep.~Electrical \& Computer Eng., Seoul National University, \email{{\tt kjs9044@snu.ac.kr}, {\tt hshim@snu.ac.kr}}
	\and Kyoohyung Han \at Dep.~Mathematical Sciences, Seoul National University, \email{\tt satanigh@snu.ac.kr}}
	
	\maketitle

	\abstract{The cryptosystem based on the Learning-with-Errors (LWE) problem is considered as a post-quantum cryptosystem, because it is not based on the factoring problem with large primes which is easily solved by a quantum computer. Moreover, the LWE-based cryptosystem allows fully homomorphic arithmetics so that two encrypted variables can be added and multiplied without decrypting them. This chapter provides a comprehensive introduction to the LWE-based cryptosystem with examples. A key to the security of the LWE-based cryptosystem is the injection of random errors in the ciphertexts, which however hinders unlimited recursive operation of homomorphic arithmetics on ciphertexts due to the growth of the error. We show that this limitation can be overcome when the cryptosystem is used for a dynamic feedback controller that guarantees stability of the closed-loop system. Finally, we illustrate through MATLAB codes how the LWE-based cryptosystem can be customized to build a secure feedback control system. This chapter is written for the control engineers who do not have background on cryptosystems.}
	\abstract*{The cryptosystem based on the Learning-with-Errors (LWE) problem is considered as a post-quantum cryptosystem, because it is not based on the factoring problem with large primes which is easily solved by a quantum computer. Moreover, the LWE-based cryptosystem allows fully homomorphic arithmetics so that two encrypted variables can be added and multiplied without decrypting them. This chapter provides a comprehensive introduction to the LWE-based cryptosystem with examples. A key to the security of the LWE-based cryptosystem is the injection of random errors in the ciphertexts, which however hinders unlimited recursive operation of homomorphic arithmetics on ciphertexts due to the growth of the error. We show that this limitation can be overcome when the cryptosystem is used for a dynamic feedback controller that guarantees stability of the closed-loop system. Finally, we illustrate through MATLAB codes how the LWE-based cryptosystem can be customized to build a secure feedback control system. This chapter is written for the control engineers who do not have background on cryptosystems.}

	\section{Introduction}\label{sec:intro}

	Applications of homomorphic cryptography to the feedback controller are relatively new.
	To the authors' knowledge, the first contribution was made by Kogiso and Fujita \cite{2015-cdc-kosigo} in 2015, followed by Farokhi {\it et al.}~\cite{2016-necsys-farokhi} and Kim {\it et al.}~\cite{2016-necsys-kim} both in 2016.
	Interestingly, each of them uses different homomorphic encryption schemes; \mbox{El-Gamal} \cite{Ell85}, Paillier \cite{Pai99}, and LWE \cite{Gen09} are employed, respectively.
	{\color{orange}Because other two schemes are introduced in other chapters in this book,} this chapter is written for introducing the LWE-based cryptosystem and its customization for building a {\em dynamic} feedback controller.
	
\begin{figure}[t]
		\centering
		\includegraphics[width=.6\textwidth]{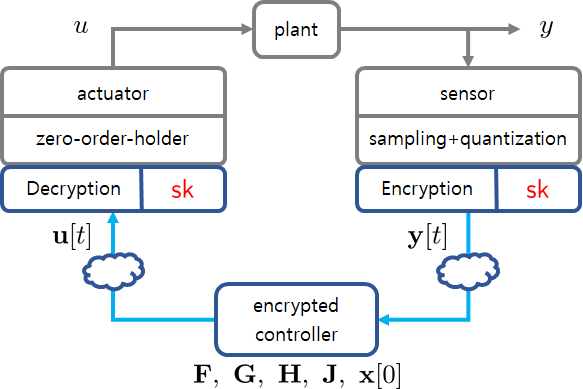}
		\caption{The control system configuration considered in this chapter. Note that the secret key ${\sf sk}$ is kept only in the plant side, and there is no need to store ${\sf sk}$ beyond the network. The parameters of the controllers (such as ${\bf F}$, ${\bf G}$, ${\bf H}$, and ${\bf J}$, as well as the initial condition ${\bf x}[0]$) are also encrypted. In this chapter, bold fonts imply encrypted variables.}
		\label{fig:1}
	\end{figure}

	Homomorphic encryption implies a cryptographic scheme in which arithmetic operations can be performed directly on the encrypted data (i.e., ciphertexts) without decrypting them.
	When applied to the control systems, security increases because there is no need to keep the secret key inside the controller (see Fig.~\ref{fig:1}), which is supposed to be a vulnerable point in the feedback loop.
	After the idea of homomorphic encryption appeared in 1978 by Rivest {\it el al.} \cite{rivest1978data}, two {\em semi-homomorphic} encryption schemes were developed.
	One is the multiplicatively homomorphic scheme by El-Gamal \cite{Ell85} developed in 1985, and the other is the additively homomorphic scheme by Paillier \cite{Pai99} developed in 1999.
	Homomorphic encryption schemes that allow both addition and multiplication appeared around 2000, and they are called {\em somewhat-homomorphic} because, even if both arithmetics are enabled, the arithmetic operations can be performed only finite times on an encrypted variable.
	In 2009, Gentry \cite{Gen09} developed an algorithm called `bootstrapping' which finally overcame the restriction of finite number of operations.
	By performing the bootstrapping regularly on the encrypted variable, the variable becomes like a newborn ciphertext and so it allows more operations on it.
	The encryption scheme with this algorithm is called {\em fully homomorphic}.
	However, fully homomorphic encryption sometimes simply implies a scheme that allows both addition and multiplication, and we follow this convention.
	
	In this chapter, we introduce
	the LWE-based fully homomorphic encryption scheme.
	We illustrate that, if the scheme is used with a stable closed-loop system then, interestingly, there is no need to employ the bootstrapping for infinite number of arithmetic operations as long as the system matrix of the controller consists of integer numbers, and the actuator and the sensor sacrifice their resolutions a little bit.\footnote{If one needs to use the bootstrapping because his/her application does not satisfies these conditions, then he/she may refer to \cite{2016-necsys-kim} in which a method to orchestrate the bootstrapping and the {\em dynamic} feedback controller has been presented.}
	Moreover, by utilizing the fully homomorphic arithmetics, we are able to encrypt all the parameters in the controller, as seen in Fig.~\ref{fig:1}.

	This chapter consists of three parts.
	In the first part (Section \ref{sec:lwesection}), we present an introduction to the LWE-based encryption, discuss the homomorphic arithmetics, and illustrate the error growth in the ciphertexts.
	The second part (Section \ref{sec:controller}) is about customization of the LWE-based cryptosystem for the linear time-invariant dynamic feedback controllers.
	In the last part (Section \ref{sec:stability}), we show the error growth can be handled by the stability, so that the dynamic controller operates seamlessly with unlimited times fully homomorphic arithmetics.
	In Section \ref{sec:needforinteger}, we conclude the chapter with a discussion on the need for integer system matrix of the controller, which is related to one of future research issues. 
	For pedagogical purposes, we simplify many issues that should be considered in practice, and instead, focus on the key ingredients of homomorphic encryptions.
	In the same respect, the codes presented in this chapter consist of simple MATLAB commands that may not be used in real applications even if it works for simple examples.

	\section{Cryptosystem based on Learning-with-Errors Problem}\label{sec:lwesection}

	We first present how to encrypt and decrypt a message, in order for the reader to look and feel the ciphertexts in the LWE-based cryptosystem.
	Then, we briefly introduce the learning-with-errors (LWE) problem because the security of the presented cryptosystem is based on the hardness of this problem.
	We also explain how the homomorphic arithmetics are performed in the LWE-based cryptosystem.
	
	Now, with $p \in \N$, let the set of integers bounded by $p/2$ be denoted by
	\begin{equation}\label{eq:modspace}
	[p] := \left\{ i \in \Z : -\frac{p}{2} \le i < \frac{p}{2} \right\}
	\end{equation}
	so that the cardinality of $[p]$ is $p$.
	In addition, we need the following set.
	
	\begin{tips}{Set of integers modulo $q$}
		Let $\Z_q$ be the set of integers modulo\footnote{In this chapter, all the modulus are chosen as powers of 10 for convenience of understanding while they are often powers of 2 in practice.} $q \in \N$.
		This means that any two integers $a$ and $b$ are regarded as the same elements of $\Z_q$ if $(a-b)\text{ mod }q = 0$.\footnote{$x\text{ mod }q$ is the remainder after division of $x$ by $q$. In this chapter, we suppose the remainder is an element of $[q]$; for example, if the remainder is greater than or equal to $q/2$, make it negative by subtracting $q$,
		e.g., $17\text{ mod }10 = -3 \in [10]$.}
		By this rule, all the integers are related with other integers that have the same remainder when divided by $q$, and thus, $[q]$ can represent $\Z_q$ if any integer $a \in \Z_q$ is treated as $b \in [q]$ such that $(a-b)\text{ mod }q = 0$.
		In this sense, $\Z_q$ is closed under  addition, subtraction, and multiplication.
	\end{tips}

	\subsection{LWE-based Cryptosystem}\label{sec:cryptobasedLWE}
	
	Let the set $[p]$ be where the integer to be encrypted belongs to, and let us denote it by $\PP$ and call it by the plaintext space.
	An element $m \in \PP$ is called a message or a {\em plaintext}.
	The value of $p$ can be chosen as a power of 10 such that $|m| < p/2$ for all messages to be used.
	Now, let $\Z_q$ be a set of integers modulo $q$ where $q = Lp$ with $L$ being a power of 10.
	Finally, in order to encrypt and decrypt a message $m$, choose a {\em secret key} ${\sf sk}$ which is an integer vector of size $N$ such that
	${\sf sk} \in \Z_q^N$.
	These $L$ and $N$ are the parameters of the LWE-based cryptosystem and their selection is discussed in more detail in Section \ref{sec:parameters}.
	
	Now, let us encrypt a column vector $m \in \PP = [p]^n$ having $n$ elements in $[p]$.
	Whenever a new message $m$ is encrypted, a new random matrix
	${\sf A} \in \Z_q^{n \times N}$ is sampled from the uniform distribution over $\Z_q^{n \times N}$,
	and a new vector ${\sf e} \in [r]^n$ is randomly sampled where $r < L$, so that each component ${\sf e_{\mathnormal i}}$ satisfies that $|{\sf e_{\mathnormal i}}| < L/2$ for $i=1,\cdots,n$.
	With them, compute
	\begin{equation}\label{eq:enc}
	{\sf b} \leftarrow (- {\sf A} \cdot {\sf sk} + L m + {\sf e}) \mod q
	\end{equation}
	where $\cdot$ is the standard multiplication of a matrix and a vector.
	Then, the {\em ciphertext} ${\bf m}$ of the plaintext vector $m$ is obtained by the matrix ${\bf m} = [{\sf b}, {\sf A}] \in \CC = \Z^{n \times (N+1)}_q$ where $\CC$ is called the ciphertext space.
	Define the secret key vector ${\sf s} := [1, {\sf sk}^T]^T$ and let $\lceil \cdot \rfloor$ be the rounding operation for vectors.\footnote{We round away from zero for negative numbers, e.g., $\lceil -2.5 \rfloor = -3$ while $\lceil -2.4 \rfloor = -2$. When $a$ is a vector, $\lceil a \rfloor$ implies component-wise rounding.}
	Then, the ciphertext ${\bf m}$ is decrypted as
	\begin{equation}\label{eq:dec}
	\left\lceil \frac{({\bf m} \cdot {\sf s})  \mod q}{L} \right\rfloor  = \left\lceil \frac{Lm + {\sf e}}{L} \right\rfloor \rightarrow m
	\end{equation}
	because the size of each element of ${\sf e}$ is less than $L/2$.
	One of the key ingredients in the LWE-based scheme is the vector ${\sf e}$, which is intentionally injected in the ciphertext by \eqref{eq:enc} (and is eliminated by the decryption \eqref{eq:dec}).
	This vector is called ``error,'' and it will be seen in Section \ref{sec:lwe} that this error makes this encryption scheme secure.
	
	The discussions so far yield the MATLAB codes with an example parameter set
	\begin{verbatim}
	env.p = 1e4;  env.L = 1e4;  env.r = 1e1;  env.N = 4;
	\end{verbatim}
	which are put in a structure variable {\texttt env}.
	We randomly select a secret key by\footnote{Since the {\tt mod} function in MATLAB always returns non-negative remainders, we use our customized function (starting with the capital {\tt M})
		$$\text{\tt function y = Mod(x,p), y = mod(x,p); y = y - (y >= p/2)*p; end}$$
		in order to have signed results in the set like \eqref{eq:modspace}.}\footnote{In order to run the codes in this chapter, please choose small numbers for the secret key in order not to cause the overflow of the double variable in MATLAB. An example is to replace {\tt env.q*env.L} by 10 for example.}
	\begin{verbatim}
	sk = Mod(randi(env.q*env.L,[env.N, 1]), env.q*env.L);
	\end{verbatim}
	Under these parameters, the codes are as follows.

	\begin{programcode}{Functions $\Enc$, $\Dec$}
		\begin{verbatim}
		function ciphertext = Enc(m,sk,env)
		n = length(m);   q = env.L*env.p;
		A = randi(q, [n, env.N]);
		e = Mod(randi(env.r, [n,1]), env.r);
		b = -A*sk + env.L*m + e;
		ciphertext = Mod([b,A], q);
		end
		
		function plaintext = Dec(c,sk,env)
		s = [1; sk];
		plaintext = round( Mod(c*s, env.L*env.p)/env.L );
		end
		\end{verbatim}
	\end{programcode}
	
	An example run shows encryption and decryption of a number 30:
	\begin{verbatim}
	sk =
	-13203881	
	-22462885
	-28840424
	4713455
	
	c = Enc(30,sk,env)
	c =
	-43264645    21696438   -15923263    46660236    20129426
	
	m = Dec(c,sk,env)
	m =
	30
	\end{verbatim}
	
	It is seen from the outcome that the message ${\tt 30}L$ is hiding in the number $-{\sf A}\cdot{\sf sk} + {\sf e}$ that is the first element of ${\bf m}$.

	\subsubsection{Necessity of error injection and the learning-with-errors problem}\label{sec:lwe}
	
	A measure for the security of a cryptosystem is how hard it is to find the secret key ${\sf sk}$ when arbitrarily many pairs $(m_i, {\bf m}_i)$ are given.
	In fact, the ciphertexts ${\bf m}_i$ are easily available to the adversary by eavesdropping the communication line, and there are many cases that the plaintexts $m_i$ are also obtainable.
	(For example, one may guess an email begins with the word ``Dear'' even if it is encrypted.)
	When the pair $m_i$ and ${\bf m}_i = [{\sf b}_i, {\sf A}_i]$ is available, the adversary can easily obtain $(-{\sf A}_i \cdot {\sf sk} + {\sf e})$ as well as ${\sf A}_i$ by subtracting $L m_i$ from ${\sf b}_i$ (see \eqref{eq:enc}).
	Hence, if there is no error injection of ${\sf e}$, then the problem of searching ${\sf sk}$ simply becomes solving a linear equation in $\Z^N_q$, which is not difficult.
	
	Interestingly, with the error ${\sf e}$ injected, it was proved that solving (or, `learning') ${\sf sk}$ becomes extremely difficult.
	This problem is called `learning-with-error (LWE)' problem, which has been introduced in \cite{Regev09}.
	For example, with $\bar {\sf s} = [{\sf s_1}, {\sf s_2}, \dots, {\sf s_4}]^T = [3,-5,1,0]^T \in \Z^4_{100}$, consider a sequence of linear equations with errors:
	\begin{align*}
	32 {\sf s_1} + 17 {\sf s_2} -5 {\sf s_3} + 8 {\sf s_4} + {\sf e_1} &= 6 + {\sf e_1} = 7 & &\pmod{100} \\
	-6 {\sf s_1} + 17 {\sf s_2} + 1 {\sf s_3} + 18 {\sf s_4} + {\sf e_2} &= -2 + {\sf e_2} = -3 & &\pmod{100} \\
	44 {\sf s_1} + 32 {\sf s_2} + 12 {\sf s_3} + 28 {\sf s_4} + {\sf e_3} &= -16 + {\sf e_3} = -15 & &\pmod{100} \\
	\vdots &
	\end{align*}
	With the error (which need not be large; just a small error is enough, e.g., the error of $1, -1, 1, \cdots$ is used in the above example), finding $\bar {\sf s}$ (and ${\sf e}_i$ as well) in the set $\Z^4_{100}$ becomers harder (and it becomes very difficult when the dimension gets higher).
	In fact, this problem is known to be as hard as the worst-case ``lattice problem'' 
	so that the cryptosystem based on it becomes secure at the same level of difficulty to solve the problem.
	Actually, the cryptosystem based on the LWE problem is known to be as much secure as even the quantum computer takes long time to solve (and thus, it is known as a post-quantum cryptosystem \cite{NISTreport16}).
	This is because the difficulty is not based on the factoring problem,\footnote{Factoring problem is to find large prime numbers $p$ and $q$ when $N=pq$ is given. This problem is supposed to be easily solved (i.e., in polynomial time) by quantum computers.} which has been a basis for many other cryptosystems.

	\subsubsection{How to choose parameters for desired level of security?}\label{sec:parameters}
	
	An encryption scheme is called $\lambda$-bit secure, whose meaning is briefly introduced in this subsection.
	For this, let us consider a game between an adversary and a challenger.
	The rule is that, whenever the adversary submits two messages $m_1$ and $m_2$ to the challenger, the challenger randomly chooses one of them with equal probability and returns it back to the adversary after encrypting the chosen message. 
	Then, the adversary guesses which one of $m_1$ and $m_2$ is encrypted by inspecting the received ciphertext.
	If the ratio of the adversary's correct guess is not meaningfully greater than $0.5$ as the game repeats, then the encryption scheme is said to be {\em indistinguishable}.
	Let $\mathcal{A}$ denote the algorithm that the adversary uses in the game to guess.
	Then, the encryption scheme is called \textit{$\lambda$-bit secure} if, for all available adversary's algorithm $\mathcal{A}$, it holds that
	$$(\textsf{Computation complexity of $\mathcal{A}$}) \times \left(\dfrac{1}{\left|0.5 - \textsf{Success probability of $\mathcal{A}$}\right|}\right) > 2^\lambda.$$
	Clearly, large $\lambda$ implies that the adversary needs high computational complexity while the success rate is not very different from 0.5 for all possible attack algorithms.
	
	For the case of the LWE-based cryptosystem used in this chapter, it is rather convenient to assess its level of security by using a useful tool, called `LWE estimator.' 
	This tool is implemented using \texttt{Sage} program language and an on-line version is also available.\footnote{\tt https://bitbucket.org/malb/lwe-estimator}
	When the parameters $p$, $L$, $r$, and $N$ of the LWE-based cryptosystem is given, the estimator computes expected number of operations to attack the encryption scheme by various attack algorithms $\mathcal{A}$, and finally returns $\lambda$.
	Below is an example for using the estimator with a specific parameter set.
	
	\begin{verbatim}
	load(estimator.py)
	p = 1e4; L = 1e4; r = 1e2; N = 20
	_ = estimate_lwe(N, (r / (L * p)), (L * p))
	\end{verbatim}
	
	\noindent
	The following is the output of the estimator with the parameter set.
	
	\begin{verbatim}
	usvp: rop: = 2^29.7, ...
	dec: rop: = 2^32.3, ...
	dual: rop: = 2^31.3, ...
	\end{verbatim}
	
	\noindent
	Those values of \texttt{rop} mean the number of operations for each attack called {\tt usvp}, {\tt dec}, and {\tt dual}.
	Therefore, we can say that the parameter set has at least 29.7-bit security.
	It is known that the security level $\lambda$ roughly has the following property:
	$$\frac{N}{\log q - \log r} \propto \frac{\lambda}{\log \lambda}.$$
	Therefore, increasing $N$ may easily lead to higher security.

	\subsection{Homomorphic Property of LWE-based Cryptosystem}\label{sec:homo}

	As a control engineer, a reason for particular interest on the LWE-based cryptosystem is that it allows homomorphic arithmetics. 
	By homomorphic arithmetics, we mean, for two plaintexts $m_1$ and $m_2$, it holds that
	$$\Dec\big( \Enc(m_1) \; \ast_\CC \; \Enc(m_2) \big) = m_1 \; \ast_\PP \; m_2$$
	where $\ast_\PP$ and $\ast_\CC$ are binary operations on the plaintext space $\PP$ and the ciphertext space $\CC$, respectively, and $\Dec$ and $\Enc$ symbolize the encryption and the decryption functions, respectively.
	The LWE-based cryptosystem provides both the homomorphic addition and the homomorphic multiplication.
	
	The addition is defined in the following way:
	\begin{align}\label{eq:homoadd}
	\begin{split}
	\Enc(m_1) &+_{\CC} \Enc(m_2) = {\bf m_1} + {\bf m_2} \\
	&= \left[ - {\sf A_1} \cdot {\sf sk} + L m_1 + {\sf e_1}, \; {\sf A_1} \right] + \left[ - {\sf A_2} \cdot {\sf sk} + L m_2 + {\sf e_2}, \; {\sf A_2} \right] \\
	&= \left[ - ({\sf A_1}+{\sf A_2}) \cdot {\sf sk} + L (m_1+m_2) + ({\sf e_1}+{\sf e_2}), \; {\sf A_1}+{\sf A_2} \right]
	\end{split}
	\end{align}
	where $+_{\CC}$ is the addition on the ciphertext space and $+$ is the standard matrix addition.
	To see the homomorphic property, observe that
	\begin{equation}\label{eq:adddec}
	\Dec({\bf m_1}+{\bf m_2}) = \left\lceil \frac{({\bf m_1}+{\bf m_2})\cdot {\sf s} \mod q}{L} \right\rfloor = \left\lceil  \frac{L(m_1+m_2)+{\sf e_1}+{\sf e_2}}{L}  \right\rfloor = m_1 + m_2
	\end{equation}
	as long as $m_1+m_2\in [p]$ and	each element of $|{\sf e_1}+{\sf e_2}|$ is less than $L/2$.
	
	Let us now consider the homomorphic multiplication of two {\em scalar} plaintexts $m_1 \in [p]$ and $m_2 \in [p]$.
	Without loss of generality, let $m_1$ be the multiplicand and $m_2$ be the multiplier for the product
    $m_2m_1 $.
	For the multiplicand $m_1$, we use the previous encryption function ${\bf m_1} = \Enc(m_1) \in \Z^{1 \times (N+1)}_q$ but for the multiplier $m_2$, we slightly change the encryption method\footnote{
	In practice, parameters $d\in\N$ and $\nu\in\N$ is chosen such that $q=\nu^d$,
	which customizes
	the dimension of the ciphertext ${\bf M}_2$ as
	 ${\bf M}_2\in \Z^{d(N+1) \times (N+1)}_q$.
	}\footnote{This encryption was developed in \cite{GSW13} and the idea of using different encryption method for $m_2$ was introduced in \cite{CGG+16}, which is customized for our context in this chapter.}
	as
	\begin{equation}
	{\bf M_2} = \Encm(m_2) = m_2 R + \Enc(0_{\log q \cdot (N+1) \times 1}) \; \in \Z^{\log q\cdot(N+1) \times (N+1)}_q
	\end{equation}
	where $0_{\log q\cdot(N+1) \times 1}$ is the plaintext of zero vector in $[p]^{\log q\cdot(N+1) \times 1}$ and, with the Kronecker product being denoted by $\otimes$, 
	$$R := [10^0, 10^1, 10^2, \cdots, 10^{\log q - 1}]^T \otimes I_{N+1}$$
	where $I_{N+1}$ is the identity matrix
	of size $N+1$, so that $R$ is a matrix of $\log q \cdot (N+1)$ by $(N+1)$.
	Note that ${\bf O} := \Enc(0_{\log q\cdot(N+1) \times 1})$ is a matrix in $\Z^{\log q\cdot(N+1) \times (N+1)}_q$, each row of which is an encryption of the plaintext 0 (but they are all different due to the randomness of ${\sf A}$ and ${\sf e}$).
	This modified encryption $\Encm$ has the same level of security as $\Enc$ because the plaintext $m_2$ is still hiding in the ciphertext ${\bf O}$.
	Now we note that any vector $c$ in $\Z^{1 \times (N+1)}_q$ can be represented using the radix of 10 as $c = \sum_{i=0}^{\log q - 1} c_i \cdot 10^i$ in which each component of the row vector $c_i$ is one of the single digit $0, 1, 2, \cdots, 9$.
	Therefore, one can define the function $D:\Z^{1\times(N+1)}_q \to \Z^{1 \times \log q \cdot (N+1)}_q$ that decomposes the argument by its string of digits as
	\begin{equation}\label{eq:decomp}
	D(c) := \begin{bmatrix} c_0, & c_1, & \cdots, & c_{\log q - 1} \end{bmatrix}.
	\end{equation}
	An example when $q = 10^2$ and $N=2$ is:
	$$D\left([40, 35, -27]\right) = D\left([40, 35, 73]\right) = [ 0, 5, 3, 4, 3, 7 ]$$
	because $[40,35,73] = [0,5,3] \cdot 10^0 + [4,3,7] \cdot 10^1$.
	As a result, it follows that $c = D(c)R$ for any $c \in \Z^{1 \times (N+1)}_q$.
	Now, the multiplication of two ciphertexts ${\bf m_1} = \Enc(m_1)$ and ${\bf m_2} = \Enc(m_2)$ can be done by, with ${\bf M_2} = \Enc2(\Dec({\bf m_2}))$,
    $$
	{\bf M_2} \times_\CC {\bf m_1} := D({\bf m_1}) \cdot {\bf M_2} \quad \in \Z^{1 \times (N+1)}_q$$
	where $\cdot$ is the standard matrix multiplication.
	It should be noted that the operation $\Enc2(\Dec({\bf m_2}))$ requires the secret key ${\sf sk}$, and thus, the above operation is more suitable when the multiplier $m_2$ is encrypted as ${\bf M_2} = \Enc2(m_2)$ {\it a priori}, and used repeatedly for different ${\bf m_1}$'s (which will be the case when we construct dynamic feedback controllers).
	In this case, the product of the multiplicand ${\bf m_1}$ with the multiplier ${\bf M_2}$ is simply performed as $D({\bf m_1}) \cdot {\bf M_2}$.
	To see the homomorphic property, we first note that, with the secret key $\sf s$,
	\begin{align}\label{eq:multproperty}
	\begin{split}
	({\bf M_2 \times_\CC {\bf m_1}})\cdot {\sf s} =	D({\bf m_1})\cdot {\bf M_2}\cdot {\sf s} &= D({\bf m_1})\cdot \big( m_2 R + {\bf O} \big) \cdot {\sf s} \\
	&= m_2 {\bf m_1}\cdot{\sf s} + D({\bf m_1}) \cdot {\sf e}_{\bf M_2}
	\end{split}
	\end{align}
	where ${\sf e}_{\bf M_2} \in \Z^{\log q \cdot (N+1) \times 1}_q$ is the error vector inside the ciphertext ${\bf O}$.
	From this, we observe that multiplication by $\bf M_2$ is equivalent to multiplication by the plaintext $m_2$ plus an error.
	Then, we have the homomorphic property for multiplication as
	\begin{align}\label{eq:multdec}
	\begin{split}
	\Dec({\bf M_2 \times_\CC {\bf m_1}})&=\left\lceil \frac{D({\bf m_1}) \cdot {\bf M_2}\cdot {\sf s}\mod q}{L}\right\rfloor\\&=\left\lceil \frac{m_2(Lm_1+{\sf e}_1) + D({\bf m_1})\cdot {\sf e_{\bf M_2}} }{L} \right\rfloor = m_2 m_1,
	\end{split}
	\end{align}
	as long as $m_2m_1 \in [p]$ and
	\begin{equation}\label{eq:gsw_error}
	\left| \frac{m_2{\sf e_1}}{L} + \frac{D({\bf m_1}) \cdot {\sf e}_{\bf M_2}}{L} \right| < \frac12.
	\end{equation}
	
	A sample run and the codes for two operations are as follows:
	\begin{verbatim}
	q = env.p*env.L;
	c1 = Enc(-2,sk,env);  c2 = Enc(3,sk,env);  c2m = Enc2(3,sk,env);
	
	c_add = c1 + c2;
	c_mul = Decomp(c1, q)*c2m;
	
	Dec(c_add,sk,env) 
	ans = 
	1
	
	Dec(c_mul,sk,env)
	ans =
	-6
	\end{verbatim}
	
	\begin{programcode}{Functions ${\sf Enc2}$, ${\sf Decomp}$}
		\begin{verbatim}
		function ciphertext = Enc2(m,sk,env)
		q = env.L*env.p;  N = env.N;  lq = log10(q);
		R = kron( power(10, [0:1:lq-1]'), eye(N+1) );
		ciphertext = Mod(m*R + Enc(zeros(lq*(N+1),1), sk, env), q);
		end
		
		function strdigits = Decomp(c, q)
		lq = log10(q);
		c = mod(c, q); 
		strdigits = [];
		for i=0:lq-1,
		  Q = c - mod(c, 10^(lq-1-i));
		  strdigits = [ Q/10^(lq-1-i), strdigits ];
		  c = c - Q;
		end
		end
		\end{verbatim}
	\end{programcode}
	
	If a ciphertext ${\bf m_1} \in \Z^{1 \times (N+1)}_q$ is multiplied with a plaintext $m_2 \in [p]$, then it is simply performed by ${\bf m_1} m_2 \pmod{q}$ because this case can be considered as a repeated homomorphic addition. 
	
	Finally, we close this section by presenting a code for obtaining the product of the ciphertext of a matrix $F \in [p]^{m \times n}$ and the ciphertext of a column vector $x \in [p]^n$ that is equivalent to the ciphertext of $Fx$.
	With ${\bf F}_{i,j} = \Enc2(F_{i,j})$ where $F_{i,j}$ is the $(i,j)$-th element of $F$, and ${\bf x}_j$ being the $j$-th row of ${\bf x} = \Enc(x)$, we define the multiplication as
	\begin{equation*}
	{\bf F} \times_\CC {\bf x} := \sum_{j=1}^n\begin{bmatrix}  D({\bf x}_j) \cdot {\bf F}_{1,j}  \\  D({\bf x}_j) \cdot {\bf F}_{2,j} \\ \vdots \\  D({\bf x}_j) \cdot {\bf F}_{n,j} \end{bmatrix}
	\end{equation*}
	in which, we abuse the notation $\times_\CC$ that was defined for a scalar product.
	Then, analogously to \eqref{eq:multproperty} and \eqref{eq:multdec}, it follows that
	\begin{align}\label{eq:multmat}
	\begin{split}
	({\bf F}\times_\CC {\bf x})\cdot {\sf s} &=\sum_{j=1}^n\begin{bmatrix}  D({\bf x}_j) \cdot {\bf F}_{1,j}\cdot {\sf s} \\  D({\bf x}_j) \cdot {\bf F}_{2,j} \cdot {\sf s}\\ \vdots \\  D({\bf x}_j) \cdot {\bf F}_{n,j} \cdot {\sf s}\end{bmatrix}
	= \sum_{j=1}^n\begin{bmatrix}
	{F}_{1,j}\cdot{\bf x}_j \cdot  {\sf s} +D({\bf x}_j)\cdot {\sf e}_{{\bf F}_{1,j}}\\  {F}_{2,j}\cdot{\bf x}_j \cdot {\sf s}+D({\bf x}_j)\cdot {\sf e}_{{\bf F}_{2,j}}\\ \vdots \\  {F}_{n,j}\cdot{\bf x}_j \cdot  {\sf s}+D({\bf x}_j)\cdot {\sf e}_{{\bf F}_{n,j}}\end{bmatrix}\\
	&= F\cdot {\bf x}\cdot {\sf s}+ \sum_{j=1}^n \begin{bmatrix}
	D({\bf x}_j)\cdot {\sf e}_{{\bf F}_{1,j}}\\
	D({\bf x}_j)\cdot {\sf e}_{{\bf F}_{2,j}}\\
	\vdots\\
	D({\bf x}_j)\cdot {\sf e}_{{\bf F}_{n,j}}
	\end{bmatrix}
	=: F\cdot {\bf x}\cdot {\sf s} +\Delta({\bf F},{\bf x})
	\end{split}
	\end{align}
	where ${\sf e}_{{\bf F}_{i,j}}$ is the error inside ${\bf F}_{i,j}$, and the homomorphic property is obtained as
	\begin{align*}
	\begin{split}
	\Dec({\bf F}\times_\CC{\bf x}) &= \left\lceil \frac1L\left(F\cdot {\bf x}\cdot {\sf s} +\Delta({\bf F},{\bf x}) \mod q\right)\right\rfloor \\
	&= Fx + \left\lceil \frac1L \left( F\cdot {\sf e}_{\bf x} + \Delta({\bf F},{\bf x}) \right) \right\rfloor = Fx
	\end{split}
	\end{align*}
	as long as
	$Fx \in [p]^m$ and
	the error $|F\cdot {\sf e}_{\bf x} + \Delta({\bf F},{\bf x})|$ is less than $L/2$.
	
	Therefore, the operation
	$${\bf F} \times_\CC {\bf x} + {\bf G} \times_\CC {\bf y}$$
	where $+$ is the standard addition, corresponds to the plaintext operation $Fx+Gy$ with two matrices $F$ and $G$, and two vectors $x$ and $y$.
	An example code and a sample run are as follows.
	\begin{verbatim}
	F = [1, 2; 3, 4];
	cF = Enc2Mat(F,sk,env);
	
	x = [1;2];
	cx = Enc(x,sk,env);
	
	cFcx = MatMult(cF,cx,env);
	Dec(cFcx,sk,env)
	ans =
	5
	11
	\end{verbatim}
	
	\begin{programcode}{Functions ${\sf Enc2Mat}$, ${\sf MatMult}$}
		\begin{verbatim}
		function cA = Enc2Mat(A,sk,env)
		q = env.p*env.L;   N = env.N;   [n1,n2] = size(A);
		cA = zeros(log10(q)*(N+1), N+1, n1, n2);
		for i=1:n1,
		  for j=1:n2,
		    cA(:,:,i,j) = Enc2(A(i,j),sk,env);
		  end
		end
		end
		
		function Mm = MatMult(M,m,env)
		[n1,n2,n3,n4] = size(M);   q = env.p*env.L;
		Mm = zeros(n3,env.N+1);
		for i=1:n3,
		  for j=1:n4,
		    Mm(i,:) = ...
		    Mod(Mm(i,:) + Decomp(m(j,:),env.p*env.L) * M(:,:,i,j), q);
		  end
		end
		end
		\end{verbatim}
	\end{programcode}

	\subsubsection{Error growth problem caused by error injection}\label{sec:lifespan}
	
	As can be seen in \eqref{eq:enc}, a newborn scalar ciphertext ${\bf m}$ has its error ${\sf e}$ whose size is less than $r/2$; that is, $\|{\sf e}\|_\infty \le r/2$ where $\|\cdot\|_\infty$ is the infinity norm of a vector.
	However, the size of the error inside a ciphertext can grow as the arithmetic operations are performed on the variable.
	For example, if $\MSE({\bf m})$ measures the maximum size of the worst-case error in ${\bf m}$, then $\MSE({\bf m_1}+{\bf m_2}) = \MSE({\bf m_1}) + \MSE({\bf m_2})$ and $\MSE({\bf m_1}m_2) = m_2 \MSE({\bf m_1})$ for a plaintext $m_2$, which can be seen from \eqref{eq:homoadd}.
	The multiplication may cause more increase of the error as can be seen in \eqref{eq:gsw_error}.
	That is, the error in the product ${\bf m}_{\sf prod}$ of the multiplicand ${\bf m_1}$ and the multiplier ${\bf M_2}$ leads to $\MSE({\bf m}_{\sf prod}) = m_2 \MSE({\bf m_1}) + 9 (r/2) \log q$ because each element of ${\sf e_{\bf M_2}}$ is the error of the newborn ciphertext so that its absolute value is less than $r/2$ and the component of $D({\bf m_1})$ ranges from 0 to 9.
	It is noted that the first term $m_2 \MSE({\bf m_1})$ is natural, but the amount of the second term may always add whenever the multiplication is performed.
	
	The discussions so far show that, if the arithmetic operations are performed many times on the ciphertext, then the error may grow unbounded in the worst case, and it may damage the message in the ciphertext.
	Damage of the message happens when the size of the error in the ciphertext becomes larger than $L/2$.
	Indeed, the following example shows this phenomenon:
	\begin{verbatim}
	c = Enc(1,sk,env);

	Dec(3*c,sk,env)
	ans = 
	3

	Dec(3000*c,sk,env)
	ans = 
	2999
	\end{verbatim}
	
	The error growth problem restricts the number of consecutive arithmetic operation on the ciphertext.
	In order to overcome the restriction, the blueprint of `bootstrapping' procedure has been developed in \cite{Gen09}.
	In a nutshell, the grown-up error can be eliminated if the ciphertext is once decrypted and encrypted back again.
	The bootstrapping algorithm reduces the size of error by performing this process without the knowledge of the secret key.
	However, the complexity of the bootstrapping process hinders from being used for dynamic feedback controls.
	We discuss how to overcome this problem
	without bootstrapping in Section \ref{sec:stability}.

	\section{LWE-based Cryptosystem and Dynamic Feedback Controller}\label{sec:controller}
		
	For a comprehensive discussion with dynamic controllers, let us consider a discrete-time single-input-single-output linear time-invariant plant:
	\begin{equation}\label{eq:plant}
	x_{\sf p}[t+1] = A x_{\sf p}[t] + Bu[t], \quad
	y[t] = C x_{\sf p}[t].
	\end{equation}
	To control the plant \eqref{eq:plant}, we suppose that a discrete-time linear time-invariant dynamic feedback controller has been designed as
	\begin{subequations}\label{eq:controller}
		\begin{align}
		x[t+1] &= Fx[t] + Gy[t] \label{eq:controller1} \\
		u[t] &= Hx[t] + Jy[t] \label{eq:controller2}
		\end{align}
	\end{subequations}
	where $x \in \R^n$ is the state of the controller, $y \in \R$ is the controller input, and $u \in \R$ is the controller output.
	Note that they are real numbers in general, and not yet quantized.
	In order to implement the controller by a digital computer, we need to quantize the signals $y$, $u$, and $x$, and to use the cryptosystem for the controller, we also need to make them integer values.
	This procedure is called `quantization' in this chapter.
	
	Quantization is performed both on the sensor signal $y[t]$, on the control parameters, and finally on the actuator signal $u[t]$.
	The quantization level for $y[t]$ is often determined by the specification of the sensor under the name of {\em resolution} $R_y$.
	Therefore, we define the quantized integer value of the signal $y[t]$ as
	\begin{equation}\label{eq:quanty}
	y[t] \; \longrightarrow \; \bar y[t] := \left\lceil \frac{y[t]}{R_y} \right\rfloor .
	\end{equation}
	For example, with $R_y = 0.1$, the signal $y[t] = 12.11$ becomes $\bar y[t] = 121$.
	This procedure is performed at the sensor stage before the encryption.
	On the other hand, the matrices in \eqref{eq:controller} are composed of real numbers in general.
	These numbers should be truncated for digital implementation, but it is often the case when the significant digits of them include fractional parts.
	In this case, we can ``scale'' the controller \eqref{eq:controller} by taking advantages of the linear system.
	Before discussing the scaling, we assume that the matrix $F$ consists of integer numbers so that the scaling for $F$ is not necessary.
	This is an important restriction and we will discuss this issue in detail in Section \ref{sec:needforinteger}.
	Now, take $G = [5.19, 38]^T$ for example.
	If those numbers are to be kept up to the fraction $1/10 =: S_G$, then the quantized $G$ can be defined as $\bar G := \lceil G/S_G \rfloor$ so that $\bar G = [52, 380]^T$.
	By dividing \eqref{eq:controller1} by $R_yS_G$, we obtain the quantized equation as
	$$\frac{x[t+1]}{S_G R_y} = F \frac{x[t]}{S_G R_y} + \frac{G}{S_G} \frac{y[t]}{R_y} \quad \xrightarrow{\text{truncation}} \quad \bar x[t+1] = F \bar x[t] + \bar G \bar y[t]$$
	where $\bar x[t] := x[t]/(S_GR_y)$ which becomes integer for all $t > 0$ if the initial condition is set as $\bar x[0] = \lceil x[0]/(S_GR_y) \rfloor$.
	Since there may be still some significant fractional numbers in the matrices $H$ or $J/S_G$ in general,
	the output equation \eqref{eq:controller2} is scaled with additional scaling factor $S_{HJ}$ as
	$$\frac{u[t]}{S_{HJ} S_G R_y} = \frac{H}{S_{HJ}} \frac{x[t]}{S_GR_y} + \frac{J}{S_{HJ}S_G} \frac{y[t]}{R_y} \quad \xrightarrow{\text{truncation}} \quad \bar u[t] = \bar H \bar x[t] + \bar J \bar y[t],$$
	where $\bar H := \lceil H/S_{HJ} \rfloor$, $\bar J := \lceil J/(S_{HJ}S_G) \rfloor$, and $\bar u[t] := u[t]/(S_{HJ}S_GR_y)$.
	Therefore, the {\em quantized controller}
	\begin{subequations}\label{eq:quantcontroller}
		\begin{align}
		\bar x[t+1] &= F \bar x[t] + \bar G \bar y[t] \label{eq:quantcontroller1} \\
		\bar u[t] &= \bar H \bar x[t] + \bar J \bar y[t] \label{eq:quantcontroller2}
		\end{align}
	\end{subequations}
	is composed of integer values, and the state $\bar x[t]$ evolves on the integer state-space.
	Finally, the real number input $u[t]$ is obtained by
	\begin{equation}\label{eq:quantu}
	\bar u[t] \quad \longrightarrow \quad u[t] = R_u \left\lceil \frac{R_y S_G S_{HJ}}{R_u} \bar u[t] \right\rfloor
	\end{equation}
	at the actuator stage, where $R_u$ is the resolution of the actuator.
	If $R_yS_GS_{HJ}/R_u$ is an integer then the rounding doesn't work because $\bar u[t]$ is integer. 
	It is clear that the digital implementation of \eqref{eq:controller}, given by \eqref{eq:quanty}, \eqref{eq:quantcontroller}, and \eqref{eq:quantu}, works well if the truncation error is small.

	Since the quantized controller \eqref{eq:quantcontroller} consists of all the integer matrices and vectors, it is straightforward to convert it to the homomorphically encrypted controller
	\begin{subequations}\label{eq:homocontroller}
		\begin{align}
		{\bf x}[t+1] &= {\bf F} \times_\CC {\bf x}[t] + {\bf G} \times_\CC {\bf y}[t] \label{eq:homocontroller1} \\
		{\bf u}[t] &= {\bf H} \times_\CC {\bf x}[t] + {\bf J} \times_\CC {\bf y}[t] \label{eq:homocontroller2}
		\end{align}
	\end{subequations}
	where
	the operations on the ciphertexts should be understood as explained in Section \ref{sec:homo}.
	Note that ${\bf y}[t] = \Enc(y[t])$ is always a newborn ciphertext for each $t$ because it is encrypted and transmitted from the sensor stage.
	Moreover, the ciphertexts ${\bf F}$, ${\bf G}$, ${\bf H}$, and ${\bf J}$ can be considered as all newborn ciphertexts
	because they are generated when the controller is set and not updated by the control operation.
	The equation \eqref{eq:homocontroller} is solved at each time step with the initial condition ${\bf x}[0] = \Enc(\bar x[0])$. 
	Under this setting, two new ciphertexts ${\bf x}[t+1]$ and ${\bf u}[t]$ are created at each time step, or the system \eqref{eq:homocontroller} is considered to be driven by ${\bf y}[t]$ with ${\bf x}[0]$.
	The vector ${\bf x}[0]$ also has the newborn error, but the error in ${\bf x}[t]$ may grow as time goes on because of the recursion in \eqref{eq:homocontroller1}.

	As an example, consider a first-order plant given by $x_{\sf p}[t+1] = \sqrt{2} x_{\sf p}[t] + u[t]$ and $y[t] = x_{\sf p}[t]$, for which a first-order dynamic feedback controller $x[t+1] = -1 \cdot x[t] + 1 \cdot y[t]$ and $u[t] = -1.414 \cdot x[t] + 0 \cdot y[t]$ stabilizes the closed-loop system.
	With the parameters $R_y = 10^{-3}$, $S_G = 1$, $S_{HJ} = 10^{-3}$,
	and $R_u =R_y S_G S_{HJ} = 10^{-6}$
	the simulation can be done for {\tt timesteps = 150} as follows.\footnote{An example parameter set for this example is ${\tt env.p = 1e9}$, ${\tt env.L = 100}$, and ${\tt env.r = 10}$.}
	
	\begin{verbatim}
	A = sqrt(2); B = 1; C = 1;        % plant
	F = -1; G = 1; H = -1.414; J = 0; % controller
	Ry = 1e-3; Sg = 1e0; Shj = 1e-3;    
	G_ = round(G/Sg); H_ = round(H/Shj); J_ = round(J/(Sg*Shj));
	cFG = Enc2Mat([F,G_],sk,env);  cHJ = Enc2Mat([H_,J_],sk,env);
	xp = -3.4;  x = 4.3;              % i.c. of plant and ctr  
	cx = Enc(round(x/(Ry*Sg)), sk, env);
	
	for i = 1:timesteps
	 y  = C*xp;                       % Plant output
	 cy = Enc(round(y/Ry), sk, env);  % Encryption
	 cu = MatMult(cHJ, [cx;cy], env); % Controller output
	 u  = Ry*Sg*Shj*Dec(cu, sk, env); % Plant input after Dec
	 xp = A*xp + B*u;                 % Plant update
	 cx = MatMult(cFG, [cx;cy], env); % Controller update
	end
	\end{verbatim}

	\section{Controlled Error Growth by Closed-loop Stability}\label{sec:stability}

	As mentioned previously, the growth of the error in the ciphertext ${\bf x}[t]$ is of major concern in this section.
	We have to suppress its growth not to go unbounded.

		Actually, the source of error growth is the arithmetic operations in \eqref{eq:homocontroller}.
		To see both the message and the error in the state ${\bf x}[t]$, let us decrypt the dynamics \eqref{eq:homocontroller} with the secret key $\sf s$ except the rounding operation; i.e.,
		we define $$\xi[t] := \frac{({\bf x}[t]\cdot {\sf s})\mod q}{L}\in\R^n$$ so that $\Dec({\bf x}[t])=\lceil\xi[t] \rfloor$, and see the evolution of $\xi$-system over real-valued signals, which in turn is equivalent to the operation of \eqref{eq:homocontroller}.
		According to the homomorphic property \eqref{eq:multmat},
		the $\xi$-system is derived as
		\begin{align}\label{eq:encryptedcontroller_real}
		\begin{split}
		\xi[t+1] &= F\xi[t] + \bar G\left( \bar y[t] + \frac{{\sf e}_{{\bf y}[t]}}{L}\right) + \frac{\Delta({\bf F},{\bf x}[t])}{L}+\frac{\Delta({\bf G},{\bf y}[t])}{L}\\
		&=: F\xi[t] + \bar G \bar y[t] +\Delta_1[t],\qquad \quad 
		\xi[0] = \bar x[0] + \frac{{\sf e}_{{\bf x}[0]}}{L},
		\\
		\bar u'[t] &= \bar H \xi[t] + \bar J \left(\bar y[t] + \frac{{\sf e}_{{\bf y}[t]}}{L} \right) + \frac{\Delta({\bf H},{\bf x}[t])}{L}+\frac{\Delta({\bf J},{\bf y}[t])}{L}\\
		&=: \bar H \xi[t] + \bar J \bar y[t] + \Delta_2[t],
		\end{split}
		\end{align}
		in which
		${\sf e}_{{\bf y}[t]}$ and ${\sf e}_{{\bf x}[0]}$ are the errors injected to the encryptions ${\bf y}[t]$ and ${\bf x}[0]$, respectively,
		$\Delta({\bf F},{\bf x}[t])$, $\Delta({\bf G},{\bf y}[t])$, $\Delta({\bf H},{\bf x}[t])$, and $\Delta({\bf J},{\bf y}[t])$ are the errors caused by ciphertext multiplication,
		which are defined as the same as in \eqref{eq:multmat},
		and $\bar u'[t]$ is defined as $\bar u'[t]:= ({\bf u}[t]\cdot {\sf s}\mod q)/L$ so that $\Dec({\bf u}[t])=\lceil \bar u'[t] \rfloor$.
		
\begin{figure}[t]
			\centering
			\includegraphics[width=.6\textwidth]{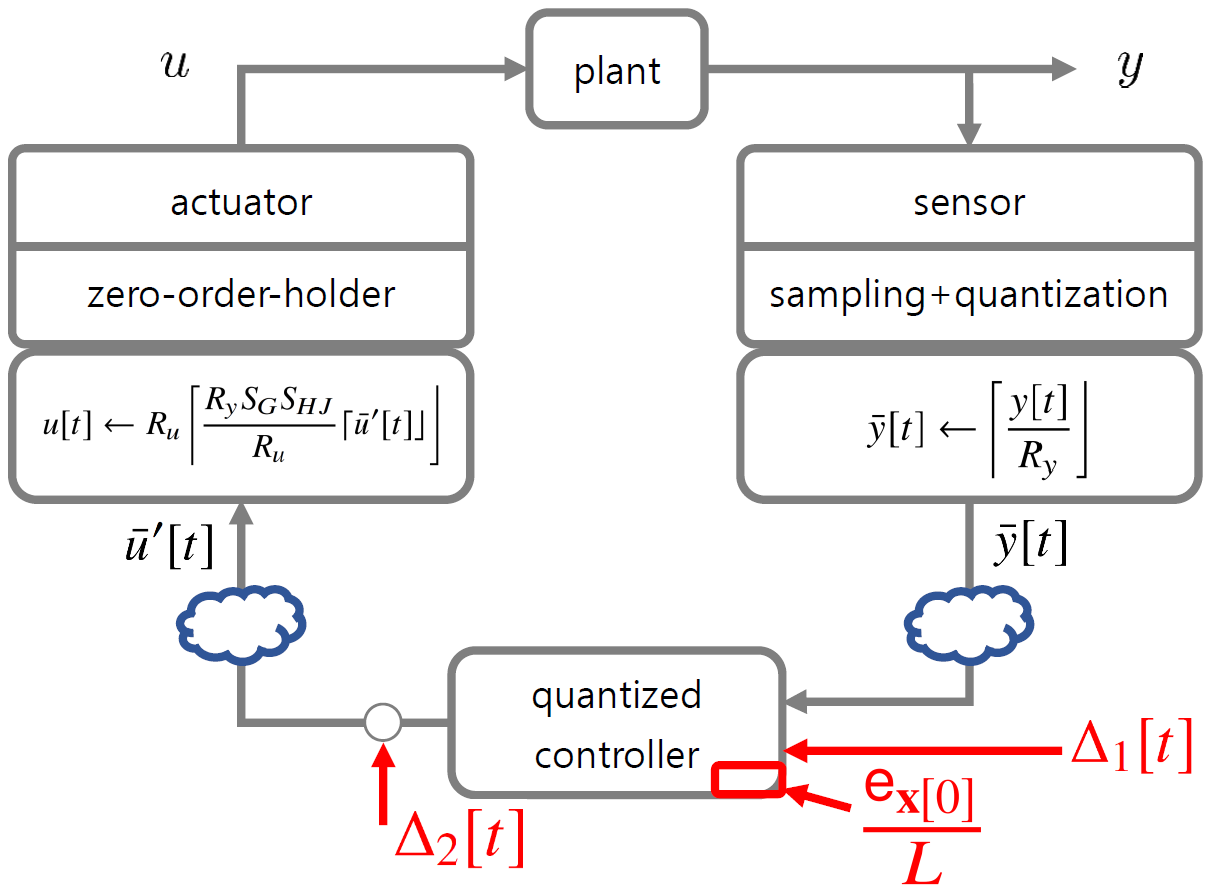}
			\caption{This figure describes the closed-loop system with the controller \eqref{eq:encryptedcontroller_real}.
			In other words, the behavior of the closed-loop system with the encrypted controller \eqref{eq:homocontroller} is the behavior of the closed-loop with the quantized controller \eqref{eq:quantcontroller} with the norm-bounded external disturbances $\Delta_1[t]$ and $\Delta_2[t]$, and the perturbation ${\sf e}_{{\bf x}[0]}/L$ on the initial condition of the controller.}
			\label{fig:2}
		\end{figure}

		For the comparison with the quantized controller \eqref{eq:quantcontroller},
		the first observation is that if there is no error injected to ciphertexts ${\bf y}[t]$, ${\bf x}[0]$, and $\{{\bf F}, {\bf G}, {\bf H}, {\bf J}\}$ so that $\Delta_1[t]$, $\Delta_2[t]$, and ${\sf e}_{{\bf x}[0]}$ are all zero,
		the operation of \eqref{eq:encryptedcontroller_real} is exactly the same way as the operation of \eqref{eq:quantcontroller}.
		Then,
		with the control perspective,
		the signals $\Delta_1[t]$ and $\Delta_2[t]$ can be understood as {\it external disturbances} injected to the feedback loop, and the quantity ${\sf e}_{{\bf x}[0]}$ can be regarded as {\it perturbation} of the initial condition (see Fig.~\ref{fig:2}).
		Here,
		the sizes of $\Delta_1[t]$, $\Delta_2[t]$, and ${\sf e}_{{\bf x}[0]}$ can be made arbitrarily small by increasing the parameter $L$ for the encryption.
		This is because $\|{\sf e}_{{\bf y}[t]}/L\|_\infty$ and $\|{\sf e}_{{\bf x}[0]}/L\|_\infty$ are less than $r/(2L)$,
		and the disturbance caused by multiplication of $\{{\bf x}[t], {\bf y}[t]\}$ by $\{{\bf F}, {\bf G}, {\bf H}, {\bf J}\}$ can also be made arbitrary small with the choice of $L$; for example, as seen in \eqref{eq:multmat}, the size of signal $\Delta({\bf F},{\bf x}[t])/L$ is bounded as
		$$\left\|\frac{\Delta({\bf F},{\bf x}[t])}{L}  \right\|_\infty \le \frac1L \sum_{j=1}^n  \left\|  \begin{bmatrix}
		D({\bf x}_j[t])\cdot {\sf e}_{{\bf F}_{1,j}}\\
		D({\bf x}_j[t])\cdot {\sf e}_{{\bf F}_{2,j}}\\
		\vdots\\
		D({\bf x}_j[t])\cdot {\sf e}_{{\bf F}_{n,j}}
		\end{bmatrix}\right\|_\infty \le \frac{9 n r\log q}{2L} = \frac{9 n r(\log p+\log L)}{2L}. $$

\begin{figure}[t]
			\centering
			\includegraphics[width=.7\textwidth]{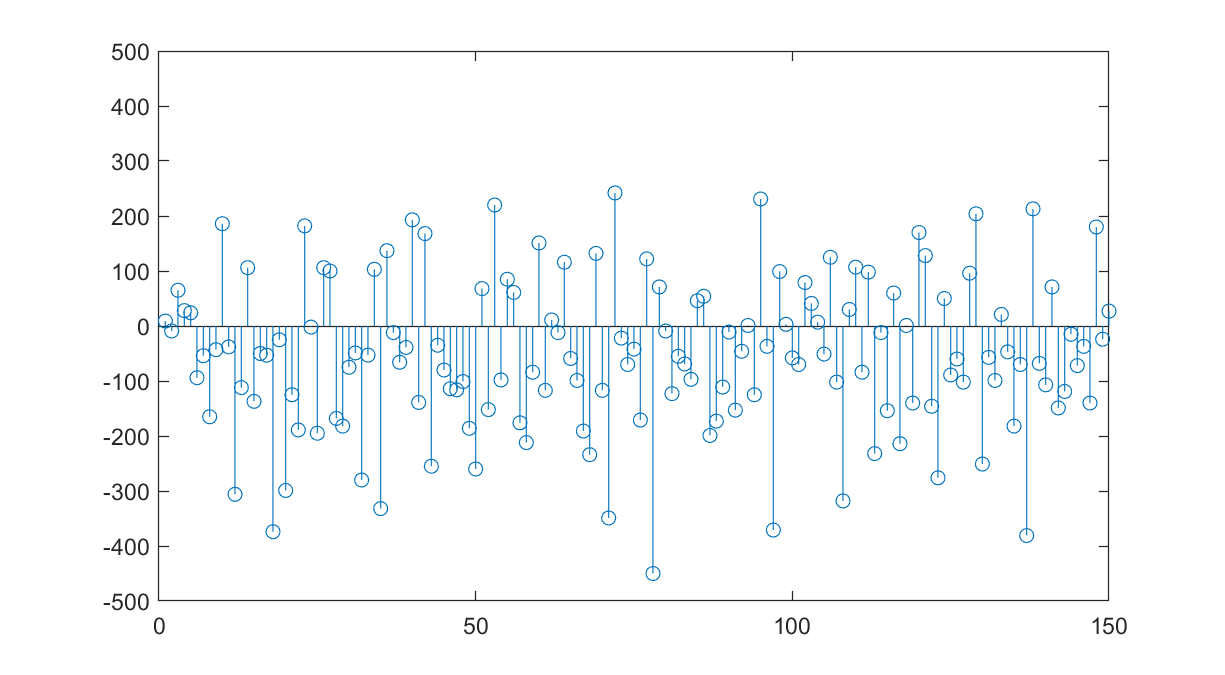}
			\caption{The error in the ciphertext ${\bf x}[t]$ from a sample run of the simulation in Section \ref{sec:controller}.
				This is the plot of $L(\xi[t] - \bar x[t])$ where $\bar x[t]$ is the state of the quantized controller and $\xi[t]$ is the state of \eqref{eq:encryptedcontroller_real}, with the parameters $L = 100$, $1/R_yS_G = 1000$, and $r = 10$.
				It is seen that the error goes beyond $L$, but is suppressed within one digit in the resolution range. This means that the message is damaged but only for one last digit.
			}
			\label{fig:3}
		\end{figure}

		Now, in terms of the error growth problem of the controller state,
		the difference $\xi[t]-\bar x[t]$ corresponds to the error of our concern.
		One might expect that
		the size of $\xi[t]-\bar x[t]$ can be made arbitrarily small by increasing $L$, but it is not true
		due to the rounding operations in the sensor and actuator;
		for example,
		if the difference $\xi[t]-\bar x[t]$ is so small that the difference of actuator inputs is less than the size of input resolution, it is truncated and the difference is not compensated in the closed-loop stability.
		As a result, the error eventually grows up to the resolution range, but is controlled not to grow more than that.
		Therefore, the damage of the message in the ciphertexts ${\bf x}[t]$ is inevitable, but it can be limited up to the last a few digits.
		Motivated by this fact, one may intentionally enhance the resolutions by a few more digits in order to preserve the significant figures.
		In this way, as long as the injected errors $\Delta_1[t]$, $\Delta_2[t]$, and ${\sf e}_{{\bf x}[0]}/L$ are sufficiently small, the error (i.e., the difference $\xi[t]-\bar x[t]$) is controlled not to grow unbounded by the closed-loop stability.
		See a simulation result in Fig.~\ref{fig:3}.
		
In the rest of this section, we analyze the control performance in terms of the encryption as well as the quantization. 
The detailed quantitative analysis is omitted but can be found in \cite{we19}.
For this, let us recall that $\Dec({\bf u}[t]) = \lceil \bar u'[t] \rfloor$.
This leads to, by \eqref{eq:quantu}, 
\begin{align*}
u &= R_u \left\lceil \frac{R_y S_G S_{HJ}}{R_u} \lceil \bar u' \rfloor \right\rfloor = R_y S_G S_{HJ} \bar u' + \Delta_{\Dec} + \Delta_u
\end{align*}
where 
\begin{align*}
\Delta_\Dec &:= R_y S_G S_{HJ} \left( \lceil \bar u' \rfloor - \bar u' \right), \quad  
\Delta_u := R_u \left\lceil \frac{R_y S_G S_{HJ}}{R_u} \lceil \bar u' \rfloor \right\rfloor - R_y S_G S_{HJ} \lceil \bar u' \rfloor
\end{align*}
in which, $\Delta_\Dec$ implies the error caused by the rounding in the decryption, and $\Delta_u$ implies the error by the quantization of the input stage.
Now, for the sake of simplicity, let us assume that $G = S_G \bar G$, $H = S_{HJ} \bar H$, and $J = S_{HJ}S_G \bar J$, which means there is no error due to the scaling of the matrices.
By defining $\xi' := R_y S_G \xi$, we obtain from \eqref{eq:encryptedcontroller_real} that  
$$R_y S_G S_{HJ} \bar u' = H \xi' + J y + J \Delta_y + R_y S_G S_{HJ} \Delta_2 \quad \text{where} \quad \Delta_y := R_y \left\lceil\frac{y}{R_y}\right\rfloor-y$$
in which, $\Delta_y$ is the error caused by the quantization at the sensor stage.
Putting together, the closed-loop system of the plant \eqref{eq:plant} and the controller \eqref{eq:encryptedcontroller_real} is equivalently described by
\begin{align*}
x_{\sf p}[t+1] &= A x_{\sf p}[t] + B(H\xi'[t] + J C x_{\sf p}[t]) \\
&\qquad\qquad\qquad + \left\{B(J\Delta_y[t] + R_yS_GS_{HJ}\Delta_2[t] + \Delta_\Dec[t] + \Delta_u[t])\right\} \\
\xi'[t+1] &= F \xi'[t] + G C x_{\sf p}[t] + \left\{ G \Delta_y[t] + S_G S_{HJ} \Delta_1[t] \right\}
\end{align*}
with the initial condition of the controller is set to be
$$\xi'[0] = x[0] + \left\{R_y S_G \left\lceil \frac{x[0]}{R_yS_G} \right\rfloor - x[0] + \frac{R_yS_G{\sf e}_{{\bf x}[0]}}{L}\right\}$$
where $x[0]$ is the initial condition of \eqref{eq:controller}.
Note that all the braced terms (i.e., errors) can be made arbitrarily small with sufficiently small $R_y$ and $R_u$ and with sufficiently large $L$.
Moreover, with all these errors being zero, the above system is nothing but the closed-loop system of the plant \eqref{eq:plant} and the controller \eqref{eq:controller}, which is supposed to be asymptotically stable.
Therefore, it is seen that the control performance with the encrypted controller \eqref{eq:homocontroller} can be made arbitrarily close to the nominal control performance with the linear controller \eqref{eq:controller}.

	\section{Conclusion and Need for Integer System Matrix}\label{sec:needforinteger}
	
	In this chapter,
	with the use of
	fully homomorphic encryption,
	we have seen a method as well as an illustrative example
	to implement a dynamic feedback controller over encrypted data.
	Exploiting both additively and multiplicatively homomorphic properties of LWE-based scheme,
	all the operations in the controller are performed over encrypted parameters and signals.
	Once the designed controller \eqref{eq:controller} is converted to the dynamical system \eqref{eq:quantcontroller} over integer,
	it can be directly encrypted as \eqref{eq:homocontroller}.
	From the nature of
	fully homomorphic encryption schemes,
	the error injected
	to the encryption ${\bf y}[t]$
	may be accumulated in the controller state ${\bf x}[t]$
	under the recursive state update and may affect the message.
	However, from the control perspective,
	it has been seen that the effect of error is controlled and suppressed by the stability of the closed-loop system.
	
	For the concluding remark,
	let us revisit that
	the encryption scheme for the dynamic controller \eqref{eq:controller}
	is based on the assumption that
	all entries of the system matrix $F$ are {\it integers}.
	To see the necessity of this assumption,
	let us suppose the matrix $F$ consists of non-integer real numbers.
	One may attempt the scaling of $F$ as $\lceil F/ S_F \rfloor$ with the scaling factor
	$1/S_F>1$
	in order to keep the fractional part of $F$,
	but this scaling is hopeless
	because it results in
	recursive multiplication by $1/S_F$ for each update of the controller.
	Indeed,
	for this case,
	it can be checked that
	the state $\bar x[t]$ of the quantized controller \eqref{eq:quantcontroller1} is multiplied by $\lceil F/S_F \rfloor$ (instead of $F$) for each time step,
	so \eqref{eq:quantcontroller1} should be remodeled as the form
	\begin{equation}\label{eq:quantcontroller_hopeless}
	\bar x[t+1] = \left\lceil \frac{F}{S_F}\right\rfloor \bar x[t] + \left\lceil \frac{G}{S_F^{t+1} S_G} \right\rfloor \bar y[t]
	\end{equation}
	with the relation $\bar x[t] = x[t]/(S_F^{t} S_G R_y)$.
	However,
	encryption of \eqref{eq:quantcontroller_hopeless} is hopeless, because in this case
	the message of the encrypted state is unbounded due to the term $1/S_F^t$.
	It will lose its value when it eventually go beyond the bound $ \pm p/2$ of the plaintext space $[p]$ represented as \eqref{eq:modspace},
	unless the state is reset to eliminate the accumulated scaling factor.
	
	This problem, which is from the constraint that encrypted variables can be multiplied by scaled real numbers only a finite number of times, is in fact one of the main difficulties of encrypting dynamic controllers having non-integer system matrix\footnote{The problem is the same for encrypted controllers based on additively homomorphic encryption schemes. See \cite{Cheon18} for the details.}.
	In this respect, one may find potential benefits of using proportional-integral-derivative (PID) controllers or finite-impulse-response (FIR) filters for the design of encrypted control system, because they can be realized with the matrix $F$ being integer as follows:
	\begin{itemize}
		\item Given an FIR filter written as $C(z) = \sum_{i=0}^n b_{n-i}z^{-i}$, and the dynamic feedback controller can be realized as
		\begin{align*}
		x[t+1] &= \begin{bmatrix}
		0 & \cdots & 0 & 0\\
		1 & \cdots & 0 & 0\\
		\vdots & \ddots & \vdots & \vdots\\
		0 & \cdots & 1 & 0
		\end{bmatrix}x[t] + \begin{bmatrix}
		1\\0\\\vdots \\ 0
		\end{bmatrix} y[t], \qquad 
		u[t] = \begin{bmatrix}
		b_{n-1}&\cdots b_1 & b_0
		\end{bmatrix}x[t] + b_n y[t].
		\end{align*}
		
		\item A discrete PID controller in the parallel form is given by
		$$ C(z) = k_p + \frac{k_i T_s}{z-1} + \frac{k_d}{\frac{T_s}{N_d} + \frac{T_s}{z-1}}$$
		where $k_p$, $k_i$, and $k_d$ are the proportional, integral, and derivative gains, respectively, 	$T_s$ is the sampling time, and $N_d \in\N$ is the parameter for the derivative filter.
		This controller can be realized as
		\begin{align*}
		x[t+1] &= \begin{bmatrix}
		2-N_d & N_d -1 \\
		1 & 0 
		\end{bmatrix}x[t] + \begin{bmatrix}
		1\\0
		\end{bmatrix} y[t], \qquad
		u[t] = \begin{bmatrix}
		b_1 & b_0
		\end{bmatrix} x[t] + b_2 y[t]
		\end{align*}
		where $b_1 = k_i T_s - k_d N_d^2/T_s$, $b_0 = k_i T_s N_d - k_i T_s + k_d N_d^2/T_s$, and $b_2 = k_p + k_d N_d/T_s$.
	\end{itemize}
	
	Another idea of approximating the effect of non-integer real numbers of $F$ has been presented in \cite{Cheon18} by using stable pole-zero cancellation.
	However, it was done at the cost of increased steady-state error in control performance.
	Further research is called for in this direction.

	\section*{Acknowledgement}
	
	The authors are grateful to Prof.~Jung Hee Cheon and Dr.~Yongsoo Song, Department of Mathematical Sciences, Seoul National University, for helpful discussions. 
	This work was supported by National Research Foundation of Korea (NRF) grant funded by the Korea government (Ministry of Science and ICT) (No. NRF-2017R1E1A1A03070342).

\end{document}